# Electrical and Thermal Transport of Layered Bismuth-sulfide EuBiS$_2$F at Temperatures between 300 and 623 K

Yosuke Goto[1,†], Joe Kajitani[2], Yoshikazu Mizuguchi[2], Yoichi Kamihara[1], and Masanori Matoba[1,*]

[1]Department of Applied Physics and Physico-Informatics, Faculty of Science and Technology, Keio University, Yokohama 223-8522, Japan

[2]Department of Electrical and Electronic Engineering, Tokyo Metropolitan University, Hachioji 192-0397, Japan

We demonstrate the electrical and thermal transport of the layered bismuth-based sulfide EuBiS$_2$F from 300 to 623 K. Although significant hybridization between Eu 4$f$ and Bi 6$p$ electrons was reported previously, the carrier transport of the compound is similar to that of F-doped LaBiS$_2$O, at least above 300 K. The lattice thermal conductivity is lower than that of isostructural SrBiS$_2$F, which is attributed to the heavier atomic mass of Eu ions than that of Sr ions.

Searching for novel thermoelectric materials is the most fundamental issue for the development of thermoelectrics because the maximum conversion efficiency of a thermoelectric device is primarily determined by the material's dimensionless figure of merit, $ZT = S^2 T \rho^{-1} \kappa^{-1}$, where $T$, $S$, $\rho$, and $\kappa$ denote the temperature, Seebeck coefficient, electrical resistivity, and thermal conductivity, respectively.[1] Bismuth telluride (Bi$_2$Te$_3$) and its related compounds have been the state-of-the-art thermoelectric materials at around room temperature since the 1950s with a $ZT$ value as high as 1.4 in a nanocrystalline bulk specimen.[2] Compared with tellurides and selenides, little attention has been paid to the thermoelectric properties of bismuth-based sulfides because of their low $ZT$ value. However, recent studies on sulfide systems have clearly shown that sulfides also exhibit attractive thermoelectric properties.[3] Aside from thermoelectricity, the superconductivity of layered bismuth sulfides, such as Bi$_4$S$_3$O$_4$[4] and LaBiS$_2$O$_{1-x}$F$_x$[5], has led to novel studies on superconductivity (the chemical formula is written according to standard nomenclature[6]). The crystallographic structures of these compounds are basically composed of alternate stacks of BiS$_2$ and oxide/fluoride layers, as shown in the inset of Fig. 1. The conduction band near the Fermi level is primarily composed of in-plane Bi 6$p$ orbitals. The thermoelectric properties of LaBiS$_2$O are suppressed by F substitution,[7] whereas they are improved by Se substitution.[8]

Recently, EuBiS$_2$F has been reported as a possible compound that exhibits charge-density-wave-like order below 280 K and superconductivity below 0.3 K without element substitution.[9] Specific heat measurement showed significant hybridization between Eu 4$f$ and Bi 6$p$ electrons, while density functional theory (DFT) calculation indicates that the valence band maximum consists of Eu 4$f$ orbitals.[9] Given the results shown in the previous study, it was controversial from the viewpoint of thermoelectricity, whether Eu 4$f$ and Bi 6$p$ hybridized orbitals are effective in modulating the Fermi surface and enhancing $S$,[10] or reducing $S$ owing to the coexistence of electrons and holes, *i.e.*, mixed conduction. In this study, we demonstrate the electrical and thermal transport of EuBiS$_2$F at temperatures between 300 and 623 K.

Polycrystalline EuBiS$_2$F was prepared by a solid-state reaction using a sealed silica tube, following the method reported by Zhai et al.[9] The relative density of the sample was calculated to be 90%. The sample purity was examined by X-ray diffraction (XRD) collected using CuK$\alpha$ radiation (Rigaku RINT 2500). Rietveld analysis was performed using the RIETAN-FP code.[11] The Hall coefficient ($R_H$) at room temperature was measured using the four-probe geometry under magnetic fields ($H$) of up to $\mu_0 H = \pm 1$ Tesla. The measurements of $\rho$, $S$, and $\kappa$ were conducted between room temperature and about 623 K using a lamp heating unit (Ulvac, MILA-5000), which was described elsewhere.[12] Measurements of the electrical and thermal transport of the polycrystalline sample were performed in the same direction.

Figure 1 shows the XRD pattern of the sample and the results of Rietveld refinement. Almost all diffraction peaks corresponded to those of the EuBiS$_2$F phase, indicating that this phase was dominant in the samples. Although diffraction peaks due to unknown impurities were also observed, the diffraction intensities of the impurity phases relative to those of EuBiS$_2$F were ~1%, suggesting that the amount of impurities was at these levels in the sample. The bond valence sum (BVS) of Eu was calculated to be 2.16(1),[13,14] which was consistent with the previously reported value,[9] indicating electron doping into the BiS$_2$ conduction layer. The negative polarity of $R_H$ at 300 K confirms that the dominant carriers are electrons at this temperature. $n_H$ was calculated to be 3.2(2) ×10$^{21}$ cm$^{-3}$ assuming a single-band model.

Figure 2 shows the electrical and thermal transport properties versus the temperature of EuBiS$_2$F. $\rho$ is 3.9 mΩcm at 300 K and it slightly decreases with increasing temperature, as shown in Fig. 2(a). On the other hand, Zhai *et al.* reported $\rho$ of ~2.2 mΩcm at 300 K with a positive temperature coefficient (d$\rho$/d$T$) between 300 and 350 K.[9] This different $\rho$–$T$ plot is probably due to self-doping owing to the mixed valence of Eu and/or inevitable Bi off-stoichiometry, which is previously examined using X-ray diffraction[15] and scanning tunneling microscopy.[16]

Figure 2(b) shows $S$ as a function of temperature. The absolute value of $S$ is ~32 µVK$^{-1}$ at 300 K and it increases almost linearly up to 623 K. Both $\rho$–$T$ and $S$–$T$ plots are similar to those of LaBiS$_2$O$_{1-x}$F$_x$ ($x \approx 0.5$),[7] suggesting that the charged carrier transport of EuBiS$_2$F at these temperatures is primarily determined by the carrier concentration provided by oxide/fluoride layer, not by the hybridization between Eu 4$f$ and Bi 6$p$ electrons.

Figure 2(c) shows the total thermal conductivity ($\kappa_{\text{tot}}$) versus temperature. The lattice thermal conductivity ($\kappa_\text{l}$) was calculated by subtracting the electronic thermal conductivity ($\kappa_{\text{el}}$) from $\kappa_{\text{tot}}$. $\kappa_{\text{el}}$ was obtained by the Wiedemann–Franz relation using the Lorenz number of $2.45 \times 10^{-8}$ WΩK$^{-2}$. $\kappa_\text{l}$ was 1.5–2.0 Wm$^{-1}$K$^{-1}$ at temperatures between 300 and 623 K. This value is distinctly lower than that of isostructural SrBiS$_2$F ($\kappa_\text{l} \approx 2.7$ Wm$^{-1}$K$^{-1}$),[17] which is attributed to the lower speed of sound resulting from the heavier atomic mass of the Eu ions than that of Sr ions.[18] Notably, the relative density of the sample in this study is 90%. According to the effective medium theory (EMT) by Maxwell-Garnett, a 14% reduction in $\kappa$ is expected from the sample with a relative density of 90% under the assumption that 10% of the pores are randomly distributed and separated spheres.[19] A fully dense sample will be required to elucidate the detailed thermal transport mechanism of these compounds. Furthermore, the EMT predicts an increase in 16% in $\rho$ for the sample with a relative density of 90%. The dimensionless figure of merit is calculated to be 0.02 at 623 K.

We briefly discuss a possible method of improving $ZT$ for these compounds. A typical way of enhancing $ZT$ is by tuning $n_\text{H}$. $ZT$ for state-of-the-art thermoelectric materials, such as Bi$_2$Te$_3$ and PbTe, is optimized at $n_\text{H}$ of $10^{19}$–$10^{20}$ cm$^{-3}$,[1] while $n_\text{H}$ for EuBiS$_2$F is $3.2(2) \times 10^{21}$ cm$^{-3}$, as described above. However, $ZT$ for LaBiS$_2$O, which has $n_\text{H}$ of ~$10^{19}$ cm$^{-3}$,[20] is on the same order as that for EuBiS$_2$F, suggesting that simply tuning $n_\text{H}$ is not an effective way of improving $ZT$ for these compounds. Band engineering is a more reasonable way of improving $ZT$, as demonstrated by Se substitution in LaBiS$_2$O.[8] Notably, the detailed mechanism of the enhancement of $ZT$ in Se-doped LaBiS$_2$O has not been clarified yet. DFT calculation is considered to be effective in shedding light on the issue as well as measurements of the transport properties of fully dense samples.

In summary, we demonstrated the electrical and thermal transport of the layered bismuth-based sulfide EuBiS$_2$F at temperatures between 300 and 623 K. Although significant hybridization between Eu 4$f$ and Bi 6$p$ electrons was previously reported, the charged carrier transport of EuBiS$_2$F is similar to that of F-doped LaBiS$_2$O, at least above 300 K. These results indicate that electron doping into the BiS$_2$ layer is detrimental to the thermoelectric properties of layered bismuth-chalcogenides. $\kappa_l$ is distinctly lower than that for isostructural SrBiS$_2$F, which is probably due to the heavier atomic mass of Eu ions than that of Sr ions.

**Acknowledgments**

This work was partially supported by research grants from Keio University, the Keio Leading-edge Laboratory of Science and Technology (KLL), Japan Society for the Promotion of Science (JSPS) KAKENHI Grant Numbers 25707031 and 26400337, and the Asahi Glass Foundation.

*E-mail: matobam@appi.keio.ac.jp

†Present address: Department of Chemical System Engineering, School of Engineering, The University of Tokyo, Tokyo 113-8656, Japan

**Figure captions**

Fig. 1.

(Color online) XRD pattern of EuBiS$_2$F and the results of Rietveld analysis. The arrows denote the diffraction peaks due to unknown impurities. The lattice parameters were calculated to be $a$ = 0.40488(1) nm and $c$ = 1.353240(6) nm. The reliability factor ($R_{wp}$) is 8.48%. The inset shows the crystallographic structure of EuBiS$_2$F.

Fig. 2.

(Color online) Electrical and thermal transport properties versus temperature ($T$) of EuBiS$_2$F.

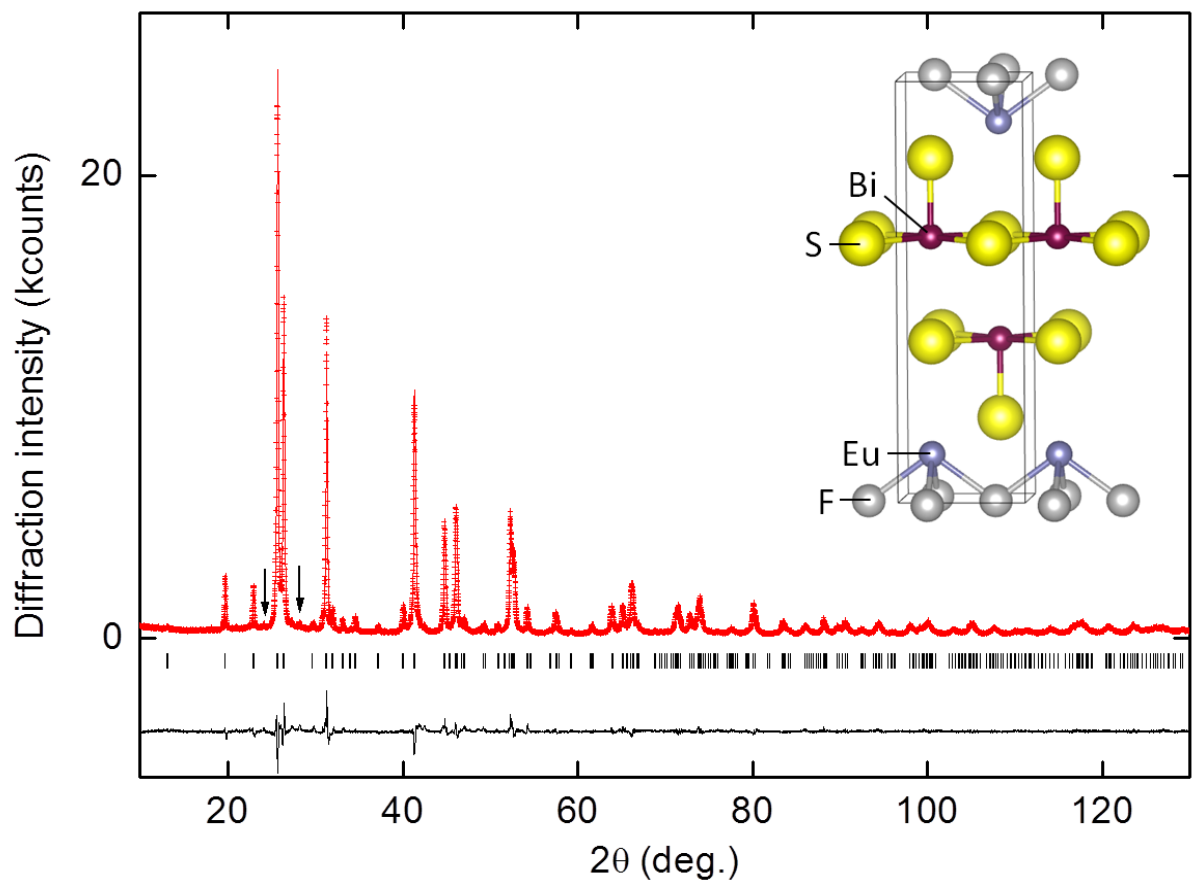

Fig. 1. (Color online)

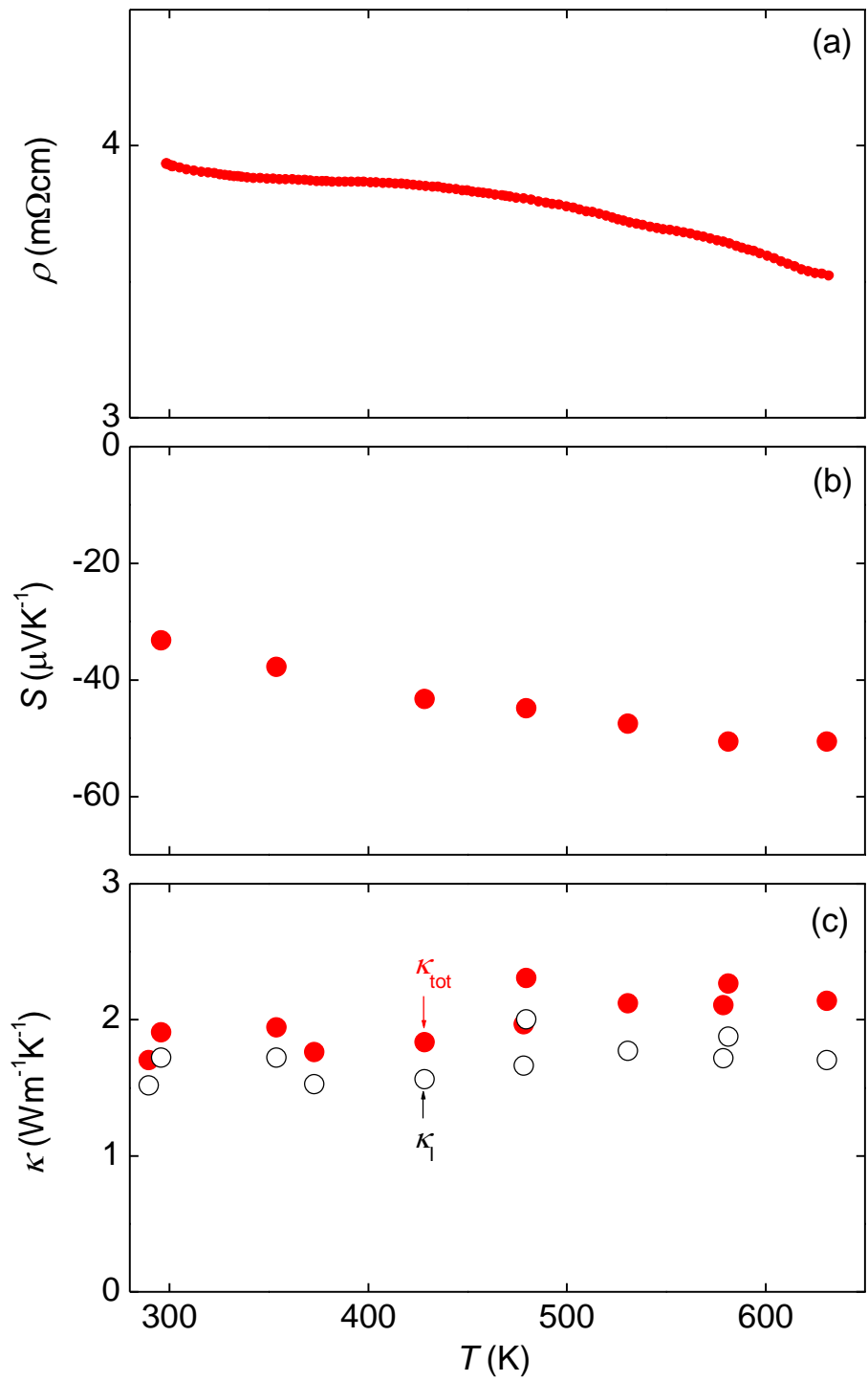

Fig. 2. (Color online)